\documentclass[12pt,twocolumn]{IEEEtran}

\IEEEoverridecommandlockouts                              
\usepackage{epsfig} 
\usepackage{amsmath} 
\usepackage{amssymb}  
\usepackage{amsfonts}

\usepackage{multirow}

\newtheorem{definition}{Definition}
\newtheorem{problem}{Problem}
\newtheorem{remark}{Remark}
\newtheorem{theorem}{Theorem}
\newtheorem{example}{Example}

\DeclareMathOperator{\ess}{ess}
\newcommand{\Registered}{\ensuremath{^{\hbox{\fontsize{5pt}{5pt}\selectfont%
      \textregistered}}}}
\title{\LARGE \bf Input-Output Finite-Time Stability}

\author{G.~De~Tommasi$^\dag$\thanks{$^\dag$ G. De Tommasi, G. Carannante and A. Pironti are with the
Dipartimento di Informatica e Sistemistica, Universit\`a degli Studi
di Napoli Federico II, Via Claudio 21, 80125 Napoli, Italy.}, R.~Ambrosino$^\ddag$\thanks{$^\ddag$ R.~Ambrosino is with Dipartimento per le Tecnologie, Universit\`{a} degli Studi di Napoli Parthenope, Napoli, Italy}, G.~Carannante$^\dag$,\\ C.~Cosentino$^\S$\thanks{$^\S$ C.~Cosentino and F.~Amato are with the School of Computer and Biomedical Engineering, Universit\`a degli Studi Magna Gr{\ae}cia di Catanzaro, Campus di Germaneto ``Salvatore Venuta'', 88100 Catanzaro, Italy.}, A.~Pironti$^\dag$, F.~Amato$^\S$}
\begin{document}

\maketitle
\thispagestyle{empty}
\pagestyle{empty}


\begin{abstract}
This paper introduces the extension of Finite-Time Stability (FTS) to the input-output case, namely the Input-Output FTS (IO-FTS). The main differences between classic IO stability and IO-FTS are that the latter involves signals defined over a finite time interval, does not necessarily require the
inputs and outputs to belong to the same class of signals, and that quantitative bounds on both inputs and outputs must be specified. This paper revises some recent results on IO-FTS, both in the context of linear systems and in the context of switching systems. In the final example the proposed methodology is used to minimize the maximum displacement and velocity of a building subject to an earthquake of given magnitude.

\medskip

\end{abstract}

\section{INTRODUCTION}\label{section:Introduction}
The concept of input-output finite time stability (IO-FTS)
has been recently introduced in~\cite{Automatica:IOFTS}; roughly speaking,
a system is said to be IO-FTS if, given a class of norm bounded input
signals defined over a specified time interval~$T$, the outputs of the system do not exceed an assigned
threshold during~$T$.

In order to correctly frame the definition of IO-FTS in the current literature, we
recall that a system is said to be IO~$\mathcal{L}_p$-stable~\cite[Ch.~5]{Khalil} if for any input of class~$\mathcal{L}_p$,
the system exhibits a corresponding output which belongs to the same class. The main differences between \emph{classic} IO stability and
IO-FTS are that the latter involves signals defined over a finite time interval, does not
necessarily require the inputs and outputs to belong to the same class, and that
\textit{quantitative} bounds on both inputs and outputs must be specified. Therefore, IO stability
and IO-FTS are independent concepts. Furthermore, while IO stability deals with the behavior of a system
within a sufficiently long (in principle infinite) time interval,
IO-FTS is a more practical concept, useful to study the behavior of the
system within a finite (possibly short) interval. Indeed, IO-FTS finds application whenever it is desired that the output variables do
not exceed a given threshold during the transients, given a certain class of
input signals.

It is important to remark that the definition of IO-FTS given in~\cite{Automatica:IOFTS}
is fully consistent with the definition of (state) FTS, where
the state of a zero-input system, rather than the input and the output, are involved.
The concept of FTS dates back to the Fifties, when it was introduced in the Russian literature
(\cite{Lebedev_jamm_1954a,Lebedev_jamm_1954b}); later during the
Sixties this concept appeared in the western control literature~\cite{Dorato:61,Michel_tct_1972}.
Recently, sufficient conditions for FTS and finite-time stabilization (the
corresponding design problem) have been provided in the control literature, see for example~\cite{Amato06,Tarb09,Shen08} in the context of
linear systems, and~\cite{Zhao08,FTSHybrid09,Du09} in the context of impulsive and hybrid systems.

\medskip

In this paper first the definition of IO-FTS is recalled, and then some recent results are revised.

\medskip

The definition of IO-FTS was originally introduced in~\cite{Automatica:IOFTS}, where two sufficient conditions to check IO-FTS when the classes of~$\mathcal L_2$ and~$\mathcal L_\infty$ inputs were considered, respectively. Both conditions required the solution of a feasibility problem involving differential linear matrix inequalities (DLMIs). In this paper we show that, in the case of~$\mathcal L_2$ signals, the condition given in~\cite{Automatica:IOFTS} is also necessary. Furthermore we also provide an alternative necessary and sufficient condition for this class of input signals, which requires that a certain Differential Lyapunov Equation (DLE) admits a positive definite solution. The efficiency
of the two conditions is discussed by means of a numerical example.

Sufficient conditions for IO-FTS of a class of hybrid systems, namely \emph{Switching Linear Systems} (SLSs), are also presented. SLSs are
linear continuous-time systems with isolated discrete switching
events, and whose state can undergo finite jump discontinuities
(\cite{Liberzon,Haddad:06}). In particular, we consider the case of \emph{time-dependent} SLSs
(TD-SLSs); for this class of switching systems the state jump and the change in the continuous
dynamic is driven by time.

\medskip

For the sake of completeness, it should be mentioned that a different concept of IO-FTS has been recently given for nonlinear systems.
In particular, the authors of~\cite{Hong:2010} consider
systems with a norm bounded input signal
over the interval~$[0\,,+\infty]$ and a nonzero initial condition. In this case, the finite-time input-output stability is related to
the property of a system to have a norm bounded output that, after a finite time interval~$T$, does not depend anymore
on the initial state. Hence, the concept of IO-FTS introduced in~\cite{Automatica:IOFTS} and the one
in~\cite{Hong:2010} are different. Note, also, that the definition of IO-FTS given in~\cite{Hong:2010} would not be well posed
in the context of linear systems, since the output of a zero-input linear system cannot go to zero in finite time. Furthermore, input-output stabilization of time-varying systems on a finite
time horizon is tackled also in~\cite{Shaked:2001}. However, as
for classic IO stability, their concept of IO stability over a
finite time horizon does not give explicit bounds on input and
output signals, and does not allow the input and output to belong
to different classes.

In~\cite{Orlov05} the problem of robust finite-time stabilization in the sense of~\cite{Bhat:00} has been framed in the context of switched systems. However, in this paper the author consider uncertainty on the nonlinear dynamic, rather than uncertainty on the resetting times set, as it is considered in this paper.

\medskip

The paper is organized as follows: in Section~\ref{section:LinearSystems} the concept of IO-FTS of linear systems is introduced, and both analysis and synthesis results are presented. Section~\ref{section:SwitchingSystems} presents result for the IO-FTS of time-dependent switching linear systems. Eventually, the design of a controller that minimizes the maximum displacement and velocity of a building subject to an earthquake of given magnitude is presented in Section~\ref{section:example}.

\medskip

\textbf{Notation.} Given a vector $v\in\mathbb{R}^n$ and a matrix~$A\in\mathbb{R}^{n\times n}$, we will denote with~$|v|$ the euclidian norm of~$v$, and with $|A|$ the induced matrix norm
\[
|A|=\sup_{v\neq0}\frac{|Av|}{|v|}\,.
\]
Given the set $\Omega~=~[t_0, t_0~+~T]$, with $t_0\in\mathbb{R}$ and $T>0$, the symbol~$\mathcal L_p(\Omega)$ denotes the space of vector-valued signals for which\footnote{For the sake of brevity, we denote by~$\mathcal{L}_p$ the set~$\mathcal{L}_p\bigl([0\,,+\infty)\bigr)$.}
\[
s(\cdot)\in\mathcal L_p(\Omega)\iff\left(\int_\Omega|s(\tau)|^p d\tau\right)^{\frac{1}{p}} <+\infty\,,
\]

Given a symmetric positive definite matrix valued function~$R(\cdot)$, bounded on $\Omega$, and a vector-valued signal~$s(\cdot)\in\mathcal L_p(\Omega)$, the weighted signal norm
\[
\left(\int_{\Omega}\bigl[s(\tau)^T R(\tau) s(\tau)\bigr]^{\frac{p}{2}} d\tau\right)^{\frac{1}{p}}\,,
\]
will be denoted by $\|s(\cdot)\|_{p\,,R}$. If~$p=\infty$
\[
\|s(\cdot)\|_{\infty\,,R} = \ess\sup_{t\in\Omega}\bigl[s^T(t)R(t)s(t)\bigr]^{\frac{1}{2}}\,.
\]
When the weighting matrix $R(\cdot)$ is constant and equal to the identity matrix~$I$, we will use the simplified notation
$\|s(\cdot)\|_{p}\,$.

\section{IO-FTS OF LINEAR SYSTEMS}\label{section:LinearSystems}
In this section we introduce the definition of IO-FTS and we revise both analysis and controller synthesis results.

\subsection{Problem Statement}\label{subsection:ProblemStatement}
In general, $w(\cdot)\in \mathcal L_p$ does not guarantee that
$y(\cdot)\in \mathcal L_p$; therefore it makes sense to give the
definition of IO $\mathcal L_p$-stability. Roughly speaking (the
precise definition is more involved, and the interested reader is
referred to~\cite[Ch.~5]{Khalil}), system \eqref{sys} is said to
be $\mathcal L_p$-stable, if $w(\cdot)\in \mathcal L_p$ implies
$y(\cdot) \in \mathcal L_p$. The
most popular cases are the ones with $p=2$ and $p=\infty$.

The concept of $\mathcal L_p$-stability is generally referred to an
infinite interval of time. In this paper we are interested to study
the input-output behavior of the system over a finite time interval.

Let us consider a linear time-varying (LTV) system in the form
\begin{subequations}\label{sys}
\begin{align}
\dot{x}(t) &= A(t)x(t)+G(t)w(t)\,,  \quad x(t_0)=0\label{sys_state}\\
y(t)  &= C(t)x(t)\,,\label{sys_out}
\end{align}
\end{subequations}
where $A(\cdot):  \Omega \mapsto \mathbb{R}^{n\times n}$, $G(\cdot):  \Omega
\mapsto \mathbb{R}^{n\times r}$, and $C(\cdot):~\Omega \mapsto \mathbb{R}^{m\times
n}$, are continuous matrix-valued functions.

\medskip

For the class of systems in the form~\eqref{sys}, let us consider the following definition.
\begin{definition}[IO-FTS of LTV systems] \label{defin:IO_FTS}
Given a positive scalar $T$, a class of input signals $\mathcal W$
defined over~$\Omega~=~[t_0~\,,~t_0~+~T]$, a continuous and positive definite
matrix-valued function~$Q(\cdot)$ defined in~$\Omega$, system~\eqref{sys} is said
to be IO-FTS with respect to $\bigl(\mathcal
W\,,Q(\cdot)\,,\Omega\bigr)$ if
\[
w(\cdot)\in \mathcal W \; \Rightarrow \;
y^T(t)Q(t)y(t)<1\,,\quad t\in \Omega\,.
\]
\hfill$\blacktriangle$
\end{definition}

\medskip

In this paper we consider the following two classes
of input signals, which will require different analysis and synthesis techniques:
\begin{itemize}
\item[i)] the set $\mathcal W$ coincides with the set of norm
bounded square integrable signals over $\Omega$, defined as
\begin{multline*}
 \mathcal{W}_2\bigl(\Omega\,,R(\cdot)\bigr):=\\ \bigl\{w(\cdot)\in \mathcal L_{2}(\Omega) \,:\,
\|w\|_{2,R}\leq 1 \bigr\} \,.
\end{multline*}
\item[ii)] \label{case 1}  The set $\mathcal W$ coincides with the set of the
uniformly bounded signals over $\Omega$, defined as
\begin{multline*}
\mathcal{W}_\infty\bigl(\Omega\,,R(\cdot)\bigr):=\\\bigl\{w(\cdot)\in \mathcal
L_{\infty}(\Omega) \,:\, \|w\|_{\infty,R}\leq 1\bigr\} .
\end{multline*}
\end{itemize}
where~$R(\cdot)$ denotes a continuous positive definite matrix-valued function. In the rest of the paper we will drop the dependency of~$\mathcal{W}$
on~$\Omega$ and~$R(\cdot)$ in order to simplify
the notation.

\medskip

Section~\ref{subsection:LinearAnalysis} provides conditions for IO-FTS when the classes of $\mathcal{W}_2$ and~$\mathcal{W}_{\infty}$ inputs are considered. These conditions are then exploited in
Section~\ref{subsection:LinearSynthesis} to solve the following design
problem, namely the problem of \emph{input-output finite-time
stabilization via dynamic output feedback}.

\medskip

\begin{problem}\label{problem:OutputFeedback}
Consider the LTV system
\begin{subequations}\label{eq:ol_system}
\begin{align}
\dot{x}(t)& = A(t)x(t)+B(t)u(t)+G(t)w(t)\,, \hspace{.2cm} x(t_0)=0 \label{eq:ol_system_a}\\
y(t)& = C(t)x(t)\label{eq:ol_system_b}
\end{align}
\end{subequations}
where $u(\cdot)$ is the control input and $w(\cdot)$ is the
exogenous input. Given the class of
signals~$\mathcal{W}$, and a continuous positive
definite matrix-valued function~$Q(\cdot)$ defined over~$\Omega$, find a dynamic output feedback controller in the form
\begin{subequations}\label{eq:feedback_sys}
\begin{align}
\dot{x_c}(t) &= A_{K}(t)x_c(t)+B_K(t)y(t)\,,\label{eq:feedback_sys_a}\\
u(t) &= C_K(t)x_c(t)+D_K(t)y(t)\label{eq:feedback_sys_b}
\end{align}
\end{subequations}
where $x_c(t)$ has the same dimension of $x(t)$, such that the
closed loop system obtained by the connection
of~\eqref{eq:ol_system} and~\eqref{eq:feedback_sys} is IO-FTS with
respect to $\bigl(\mathcal{W}\,,Q(\cdot)\,,\Omega\bigr)$. In
particular, the closed loop system is in the form
\small
\begin{subequations}\label{eq:cl_system}
\begin{align}
&
\begin{pmatrix} \dot x(t) \\ \dot x_c(t) \end{pmatrix} =
\begin{pmatrix} A+BD_KC & BC_K \\
B_K C & A_K \end{pmatrix} \begin{pmatrix}  x(t) \\ x_c(t)
\end{pmatrix}+
\begin{pmatrix} G \\ 0 \end{pmatrix} \,w(t) \nonumber \\
&  \hspace{2cm} =: A_{\mathrm{CL}}(t) \,
x_{\mathrm{CL}}(t)+G_{\mathrm{CL}}(t) \, w(t) \label{eq:cl_system_a}\\
& \hspace{.65cm} y(t) = \begin{pmatrix} C & 0 \end{pmatrix}\,
x_{\mathrm{CL}}(t)=: C_{\mathrm{CL}}(t) \,
x_{\mathrm{CL}}(t)\label{eq:cl_system_b}
\end{align}
\end {subequations}
\normalsize
where all the considered matrices depends on time,
even when not explicitly written. \hfill$\blacktriangle$
\end{problem}

\subsection{Main Results}\label{subsection:LinearAnalysis}
In this section we provide a number of conditions to check IO-FTS of linear systems. In particular, Theorem~\ref{theorem:main} provides two necessary and sufficient conditions for IO-FTS when $\mathcal{W}_2$ inputs are considered, while a sufficient condition for the class of $\mathcal{W}_{\infty}$ signals is given in Theorem~\ref{thrm:W_inf}. These results are then extended to the case of non-strictly proper systems, and to the framework of uncertain systems, respectively.

\medskip

\begin{theorem}[\cite{Automatica:IOFTS} and~\cite{Amato:CDC2011b}]\label{theorem:main}
Given system~\eqref{sys}, the class of inputs~$\mathcal{W}_2$, a continuous positive definite matrix-valued function~$Q(\cdot)$, and the time interval~$\Omega$, the following statements are equivalent:
\begin{itemize}
\item[\textbf{i)}] System~\eqref{sys} is IO-FTS with respect to~$\bigl(\mathcal{W}_2\,,Q(\cdot)\,,\Omega\bigr)$.
\item[\textbf{ii)}] The inequality
\begin{equation}\label{equation:Main_iia}
\lambda_{\max}\bigl(Q^{\frac{1}{2}}(t)C(t)W(t,t_0)C^T(t)Q^{\frac{1}{2}}(t)\Bigr)<1
\end{equation}
holds for all $t\in\Omega$, where $\lambda_{\max}(\cdot)$ denotes the maximum eigenvalue, and~$W(\cdot,\cdot)$ is the positive semidefinite solution of
\begin{subequations}\label{equation:DiffEquation}
\begin{align}
\dot{W}(t\,,t_0) &= A(t)W(t\,,t_0)+W(t\,,t_0)A^T(t)\nonumber \\&+G(t)R(t)^{-1}G^T(t)
\label{equation:DiffEquuation_1}\\
W(t_0\,,t_0) & = 0\label{equation:DiffEquuation_2}
\end{align}
\end{subequations}
\item[\textbf{iii)}]The coupled DLMI/LMI
\begin{subequations}\label{eq:DLMI_2_first}
\begin{align}
\label{eq:DLMI_2_1}& \begin{pmatrix} \dot{P}(t)+A(t)^TP(t)+P(t)A(t) & P(t)G(t)\\
G(t)^TP(t) & -R(t) \end{pmatrix} <0\\
\label{eq:DLMI_2_2} & P(t)> C(t)^TQ(t)C(t)\,,
\end{align}
\end{subequations}
admits a positive definite solution~$P(\cdot)$ over~$\Omega$.
\end{itemize}
\hfill$\blacksquare$
\end{theorem}

\medskip

\begin{remark}\label{remark:piecewiseContinuous}
Theorem~\ref{theorem:main} holds also when the system matrices in~\eqref{sys} are piecewise continuous matrix-valued functions, provided that there exists an arbitrarily small~$\epsilon>0$ such that
\[
\lambda_{\max}\bigl(Q^{\frac{1}{2}}(t)C(t)W(t,t_0)C^T(t)Q^{\frac{1}{2}}(t)\Bigr)<1-\epsilon\,,
\]
when checking~\textbf{ii)}, or that exists~$\xi>1$ such that
\[
P(t)>\xi C^T(t)Q(t)C(t)\,,
\]
when checking~\textbf{iii)}.

The case of piecewise continuous system matrices allows to give a necessary and sufficient condition for IO-FTS for the special case of switching linear systems with known resetting times, and without state jumps (see Section~\ref{IO-FTS FOR TD--IDLS/known_times}).
\hfill$\blacktriangle$
\end{remark}

\medskip

In the next example we compare the numerical efficiency when applying the two necessary and sufficient conditions of Theorem~\ref{theorem:main} to check IO-FTS of LTV systems.

\medskip

\begin{example}\label{example:analysis}
Let us consider the system
\begin{equation}\label{sys_example}
\begin{array}{ll}
A(t)=\left(\begin{array}{cc}0.5+ t & 0.1\\0.4 & -0.3+
t\end{array}\right),\,&
G=\left(\begin{array}{l}1\\1\end{array}\right),\,\\
C=\left(\begin{array}{ll}1 & 1\end{array}\right). &
\end{array}
\end{equation}
together with the following IO-FTS parameters:
\begin{equation*}\label{example_parameters}
R=1\,,\, \Omega=\bigl[0\,,0.5\bigr]\,.
\end{equation*}

The conditions stated in Theorem~\ref{theorem:main} are, in principle,
necessary and sufficient. However, due to the time-varying nature
of the involved matrices, the numerical implementation of such conditions
introduces some conservativeness.

In order to compare each other, from the
computational point of view, the
conditions stated in Theorem \ref{theorem:main}, the output weighting matrix is
left as a free parameter. More precisely, we introduce the parameter $Q_{max}$,
defined as the maximum value of the matrix $Q$ such that system \eqref{sys_example} is IO-FTS, and
use the conditions stated in  Theorem \ref{theorem:main} to obtain an estimate of $Q_{max}$.

To recast the DLMI condition~\eqref{eq:DLMI_2_first} in terms of LMIs, the matrix-valued
functions~$P(\cdot)$ has been assumed piecewise linear. In
particular, the time interval~$\Omega$ has been divided
in~$n=T/T_s$ subintervals, and the time derivatives of~$P(t)$
have been considered constant in each subinterval. It is
straightforward to recognize that such a piecewise linear function
can approximate at will a given continuous matrix function, provided that~$T_s$ is sufficiently small.

Given a piecewise linear function~$P(\cdot)$, the feasibility
problem~\eqref{eq:DLMI_2_first} has been solved by exploiting standard
optimization tools such as the Matlab LMI
Toolbox\Registered~\cite{Gahinet_matlab}.

Since the equivalence between IO-FTS and
condition~\eqref{eq:DLMI_2_first} holds when $T_s \mapsto 0$, the
maximum value of $Q$ satisfying condition~\eqref{eq:DLMI_2_first}, namely $Q_ {max}$, has
been evaluated for different values of~$T_s$. The obtained estimates
of~$Q_{max}$, the corresponding values of~$T_s$ and of the
computation time are shown in Table~\ref{table_1}. These results have been obtained by using a PC equipped with an Intel\Registered
i7-720QM processor and 4~GB of RAM.

\begin{table*}
\caption{Maximum values of $Q$ satisfying
Theorem~\ref{theorem:main} for the LTV system~\eqref{sys_example}.}
\label{table_1}
\begin{center}
\begin{tabular}{|c|c|c|c|}
\hline IO-FTS condition & Sample Time ($T_s$) & Maximum value of $Q$ & Computation time [s]\\ 

\hline \hline
\multirow{4}{*}{DLMI~\eqref{eq:DLMI_2_first}} &
0.05 & 0.2 & 2.5\\ 
\cline{2-4}
& 0.025 & 0.25 & 12.7\\
\cline{2-4}
& 0.0125 & 0.29 & 257\\
\cline{2-4}
& 0.00833 & 0.3 & 1259\\
\cline{2-4}
\hline \hline Solution of~\eqref{equation:DiffEquation} and inequality~\eqref{equation:Main_iia} & 0.003 & 0.345 & 6 \\
\hline
\end{tabular}
\end{center}
\end{table*}

We have then considered the problem of finding the maximum value of $Q$
satisfying condition~\eqref{equation:Main_iia},
where~$W(\cdot,\cdot)$ is the positive semidefinite solution
of~\eqref{equation:DiffEquation}.
In particular, equation~\eqref{equation:DiffEquation} has been
firstly integrated, with a sample time $T_s=0.003 \, s$, by using the
Euler forward method, and then the maximum value of~$Q$ satisfying
condition~\eqref{equation:Main_iia} has been evaluated by means of a linear search. As a
result, it has been found the estimate~$Q_{max}~=~0.345$, with a
computation time of about $6~s$, as it is shown in the last
row of Table~\ref{table_1}.

We can conclude that the necessary and sufficient condition based on the reachability
Gramian is much more efficient with respect to the solution of the DLMI when considering the IO-FTS analysis problem;
however, the DLMI feasibility problem is necessary in order
to solve the stabilization problem, as it is discussed in Section~\ref{subsection:LinearSynthesis}.
\hfill$\blacktriangle$
\end{example}

\medskip

\begin{theorem}[\cite{Automatica:IOFTS}]\label{thrm:W_inf} Let
$\widetilde{Q}(t)=tQ(t)$ and assume that the coupled DLMI/LMI
\begin{subequations}\label{eq:DLMI_inf}
\begin{align}
\label{eq:DLMI_inf_1}& \left(\begin{array}{cc}\dot{P}(t)+A^T(t)P(t)+P(t)A(t) & P(t)G(t)\\
G^T(t)P(t) & -R(t)\end{array}\right)<0\\
\label{eq:DLMI_inf_2} & P(t)>
C(t)^T\widetilde{Q}(t)C(t)\,,
\end{align}
\end{subequations}
admits a positive definite solution $P(\cdot)$ over~$\Omega$, then system~\eqref{sys} is IO-FTS with respect
to $\bigl(\mathcal{W}_\infty,Q(\cdot),\Omega)$. \hfill$\blacksquare$
\end{theorem}

\medskip

If system~\eqref{sys} is non-strictly-proper, i.e. if
\begin{equation}\label{equation:nonproper}
y(t)=Cx(t)+Dw(t)\,,
\end{equation}
then the following sufficient condition hold when~$\mathcal{W}_2$ signals are considered.

\begin{theorem}[\cite{Ambrosino:IFAC2011} and~\cite{Cosentino:IFAC2011}]\label{thrm:W_2}
If there exist a positive definite matrix-valued function~$P(\cdot)$ and a scalar~$\theta>1$ that solve the coupled DLMI/LMI
\begin{subequations}\label{eq:DLMI_2}
\begin{align}
\label{eq:DLMI_2_1}& \begin{pmatrix} \dot{P}(t)+A^T(t)P(t)+P(t)A(t) & P(t)G(t)\\
G^T(t)P(t) & -R(t) \end{pmatrix} <0\,,\\
\label{eq:DLMI_2_2} & \theta R(t)-R(t)> 2\theta D^T(t)Q(t)D(t)\\
\label{eq:DLMI_2_3} & P(t)> 2\theta C^T(t)Q(t)C(t)
\end{align}
\end{subequations}
over the time interval~$\Omega$, then the non-strictly-proper system~\eqref{sys_state}-\eqref{equation:nonproper} is IO-FTS with respect
to $\bigl(\mathcal{W}_2,Q(\cdot),\Omega)$.\hfill$\blacksquare$
\end{theorem}

\medskip

Starting from Theorem~\ref{thrm:W_2}, a sufficient condition for IO-FTS of non-strictly-proper systems when dealing with~$\mathcal{W}_{\infty}$ input signals can be derived letting~$\widetilde{Q}(t)=tQ(t)$, and replacing~\eqref{eq:DLMI_2_3} with
\[
P(t)\geq 2\theta C^T(t)^T\widetilde{Q}(t)C(t)\,.
\]

\medskip

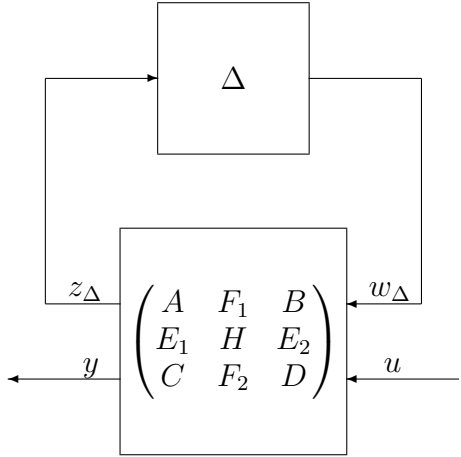
\begin{figure}
\setlength{\unitlength}{1mm}
\begin{picture}(60,60)(5,35)
\put(40,75){\framebox(20,20){$\Delta$}}
\put(25,85){\vector(1,0){15}}
\put(60,85){\line(1,0){15}}
\put(75,85){\line(0,-1){30}}
\put(75,55){\vector(-1,0){10}}
\put(68,56){$w_\Delta$}
\put(35,35){\framebox(30,30){$\begin{pmatrix}A & F_1 & B \\
                      E_1 & H & E_2 \\
                        C & F_2 & D\end{pmatrix} $}
}
\put(25,55){\line(1,0){10}}
\put(28,56){$z_\Delta$}
\put(25,55){\line(0,1){30}}
\put(80,45){\vector(-1,0){15}}
\put(70,46){$u$}
\put(35,45){\vector(-1,0){15}}
\put(30,46){$y$}
\end{picture}
\begin{center}
\caption{The uncertain system \eqref{sys_unc}.}  \label{fig_sys_unc}
\end{center}
\end{figure}
The previous results can be used to tackle the problem of IO-FTS of uncertain dynamical systems. In particular let us consider the uncertain version of system~\eqref{sys},
represented in Fig.~\ref{fig_sys_unc} and described by the following
equations
\begin{subequations} \label{sys_unc}
\begin{align}
\dot x(t)&= A(t)x(t)+F_1(t) w_\Delta(t)+B(t)u(t) \\
z_\Delta (t)&=E_1(t)x(t)+H(t)w_\Delta(t)+E_2(t) u(t) \\
y(t)&=C(t)x(t)+F_2(t)w_\Delta(t)+D(t)u(t)\\
w_\Delta(t)&=\Delta (t) z_\Delta(t) \,,
\end{align}
\end{subequations}
where $F_1(\cdot)$, $F_2(\cdot)$, $E_1(\cdot)$, and $E_2(\cdot)$
are continuous matrix-valued functions of compatible dimensions; finally
the uncertainty $\Delta(\cdot)$ is any Lebesgue measurable matrix-valued function of compatible
dimensions with $\| \Delta (t)\| \leq 1$, when~$t~\in~\Omega$.

We know extend the definition of IO-FTS to the uncertain system~\eqref{sys_unc}.
\begin{definition}[Robust IO-FTS of linear systems]
Given a interval~$\Omega$, a class of input signals $\mathcal W$
defined over $\Omega$, a positive definite matrix-valued function
$Q(\cdot)$, the uncertain system \eqref{sys_unc} is said to be
robustly IO-FTS with respect to $\bigl(\mathcal W,Q(\cdot),\Omega\bigr)$
if for any admissible uncertainty realization $\Delta (\cdot)$
the resulting linear system is IO-FTS with respect to~$\bigl(\mathcal W,Q(\cdot),\Omega\bigr)$.
\hfill$\blacktriangle$
\end{definition}

\medskip

The following theorem gives a sufficient condition for IO-FTS with respect to~$\mathcal{W}_2$. In order to avoid awkward notation, when possible we will discard the time dependence of the matrix-valued functions.
\begin{theorem}[\cite{Cosentino:IFAC2011}]\label{thrm:W_2_Robust}
If the following DLMI/LMI
\small
\begin{subequations}\label{eq:W2_Robust}
\begin{align}
\label{eq:W2_Robust_1}& \begin{pmatrix} \Psi_{11} & \Psi_{12} & \Psi_{13} & \Psi_{14}\\
\Psi^T_{12} & \Psi_{22} & 0 & \Psi_{24}\\
\Psi^T_{13} & 0 & \Psi_{33} & 0\\
\Psi^T_{14} & \Psi^T_{24} & 0 & \Psi_{44}
\end{pmatrix} <0\\
& \begin{pmatrix} 2\theta D^TQD\!\!+\!\!R\!\!-\!\!\theta R\!\!+\!\!c_3E^T_2E_2 & 2\theta D^TQF_2\!\!+\!\!c_3E^T_2H\\
2\theta F^T_2QD\!\!+\!\!c_3H^TE_2 & 2\theta F^T_2QF_2\!\!-\!\!c_3\bigl(I\!\!-\!\!H^TH\bigr)
\end{pmatrix}<0\label{eq:W2_Robust_2}\\
& \begin{pmatrix} -P\!\!+\!\!2\theta C^TQC\!\!+\!\!c_4E^T_1E_1 & 2\theta C^TQF_2\!\!+\!\!c_4E^T_1H\\
2\theta F^T_2QC\!\!+\!\!c_4H^TE_1 & 2\theta F^T_2QF_2\!\!-\!\!c_4\bigl(I\!\!-\!\!H^TH\bigr)
\end{pmatrix}<0\,,\label{eq:W2_Robust_3}
\end{align}
\end{subequations}
\normalsize
where
\begin{subequations}\label{eq:Psis}
\begin{align}
& \Psi_{11} = \dot{P}+A^TP+PA+c_1E^T_1E_1\label{eq:Psi_11}\\
& \Psi_{12} = PB\label{eq:Psi_12}\\
& \Psi_{13} = PF_1+c_1E^T_1H\label{eq:Psi_13}\\
& \Psi_{14} = PF_1\label{eq:Psi_14}\\
& \Psi_{22} = -R+c_2E^T_2E_2\label{eq:Psi_22}\\
& \Psi_{24} = c_2E^T_2H\label{eq:Psi_24}\\
& \Psi_{33} = -c_1\bigl(I-H^TH\bigr)\label{eq:Psi_33}\\
& \Psi_{44} = -c_2\bigl(I-H^TH\bigr)\label{eq:Psi_44}
\end{align}
\end{subequations}
admits a positive definite solution~$P(\cdot)$ in~$\Omega$ for a given~$c_1\,,c_2\,,c_3\,,c_4>0$, and~$\theta>1$, then system~\eqref{sys_unc} is IO-FTS with respect to $\bigl(\mathcal{W}_2,Q(\cdot),\Omega\bigr)$.
\hfill$\blacksquare$
\end{theorem}

\subsection{IO Finite-Time Stabilization via Dynamic Output Feedback}\label{subsection:LinearSynthesis}
We now exploit Theorem~\ref{theorem:main} to solve Problem~\ref{problem:OutputFeedback}. In particular, when dealing with~$\mathcal{W}_2$ signals, a necessary and sufficient condition for the IO finite-time stabilization of system~\eqref{sys} via dynamic output feedback is provided in terms of a DLMI/LMI feasibility problem.

\medskip

\begin{theorem}[\cite{Amato:CDC2011}]\label{th_of}
Given the exogenous input~$w(t)\in\mathcal{W}_2$, Problem~\ref{problem:OutputFeedback} is solvable if and only if there exist two
continuously differentiable symmetric matrix-valued functions
$S(\cdot)$, $T(\cdot)$, a nonsingular matrix-valued function
$N(\cdot)$ and matrix-valued functions $\hat A_K(\cdot)$, $\hat
B_K(\cdot)$, $\hat C_K(\cdot)$ and $D_K(\cdot)$ such that the following DLMIs are satisfied
\begin{subequations} \label{cond_fts_of}
\begin{align}
& \label{cond1_fts_of}
\begin{pmatrix}
\Theta_{11}(t) &  \Theta_{12}(t) & 0 \\
\Theta_{12}^T(t) & \Theta_{22}(t) & T(t)G(t) \\
0 & G^T(t)T(t) & -R(t)
\end{pmatrix} <0 \,, \quad t\in \Omega\\
&
\begin{pmatrix} \label{second_dmi_of}
\Xi_{11}(t) & \Xi_{12}(t) & 0\\
\Xi_{12}^T(t) & S(t)  & S(t)C^T(t)\\
0 & C(t)S(t) & Q^{-1}(t)
\end{pmatrix} > 0 \,, \quad t\in \Omega
\end{align}
\end{subequations}
where
\begin{align*}
\Theta_{11}(t) & = -\dot S(t)+A(t)S(t)+S(t)A^T(t)+B(t)\hat C_K(t)\\
&+\hat C_K^T(t)B^T(t)+G(t)R^{-1}(t)G^T(t) \\
\Theta_{12}(t) & = A(t)+\hat A_K^T(t)+B(t)D_K(t)C(t)\\
&+G(t)R^{-1}(t)G^T(t)T(t)\\
\Theta_{22}(t) & = \dot T(t)+T(t)A(t)+A^T(t)T(t)\\
&+\hat B_K(t)C(t)+C^T(t)\hat B_K^T(t) \\
\Xi_{11}(t) & = T(t)-C^T(t)Q(t)C(t) \\
\Xi_{12}(t) & = I-C^T(t)Q(t)C(t)S(t) \\
\end{align*}
\hfill$\blacksquare$
\end{theorem}

\begin{remark}[Controller design]
Assuming that the hypotheses of Theorem~\ref{th_of} are
satisfied, in order to design the controller, the following steps
have to be followed:
\begin{itemize}
\item[i)]
Find $S(\cdot)$, $T(\cdot)$,  $\hat A_K(\cdot)$, $\hat
B_K(\cdot)$, $\hat C_K(\cdot)$ and $D_K(\cdot)$ such
that~\eqref{cond_fts_of} are satisfied.
\item[ii)]
Let $M(t)=\bigl[I-T(t)S(t)\bigr]N^{-T}(t)$.
\item[iii)]
Obtain $A_K(\cdot)$, $B_K(\cdot)$ and~$C_K(\cdot)$ by inverting
\small
\begin{subequations} \label{cov_qsz_cont}
\begin{align}
& \begin{pmatrix} S(t) & I \\ I & T(t) \end{pmatrix} > 0 \label{nonsingular}\\
\hat B_K(t)&= M(t)B_K(t)+T(t)B(t)D_K(t) \\
\hat C_K(t)&=C_K(t)N^T(t)+D_K(t)C(t)S(t)\\
\hat A_K(t)&=\dot{T}(t)S(t)+\dot{M}(t)N^T(t)\nonumber\\
&+M(t)A_K(t)N^T(t)+T(t)B(t)C_K(t)N^T(t)\nonumber\\
&+M(t)B_K(t)C(t)S(t) +T(t)\bigl(A(t)\nonumber\\
&+B(t)D_K(t)C(t)\bigr)S(t) \,.
\end{align}
\end{subequations}
\normalsize
It is important to remark that, in order to invert~\eqref{cov_qsz_cont}, we need to preliminarily choose the value of~$N(t)$. The
only constraint for~$N(t)$ is to be a non singular matrix.
\end{itemize}
\hfill$\blacktriangle$
\end{remark}

\section{IO-FTS OF SWITCHING LINEAR SYSTEMS}\label{section:SwitchingSystems}
This section firstly introduces the class switching linear systems we are
dealing with in this paper, namely the time-dependent Switching
Linear Systems (TD-SLS). Sufficient conditions for the IO-FTS
of TD-SLS are presented, which require to check the
feasibility of a Difference-Differential Linear Matrix
Inequality (D/DLMI). In particular, we consider three
different cases, depending on the knowledge of the resetting times
in the time interval~$\Omega$. In Section~\ref{IO-FTS FOR TD--IDLS/known_times} we assume that the
resetting times are perfectly known. Afterwards we derive a
sufficient condition for IO-FTS when no information about the
resetting times are available, i.e. the case of arbitrary
switching signal~$\sigma(\cdot)$. Eventually, in
Section~\ref{IO-FTS FOR TD--IDLS/uncert_switching} we consider the
case of uncertain resetting times. All the results in this section are given for~$\mathcal{W}_2$ signals; however, by exploiting similar arguments as in Section~\ref{section:LinearSystems}, similar conditions can be derived when~$\mathcal{W}_\infty$ signals are considered.

Sufficient conditions for finite-time stabilization of SLS via static output feedback when the resetting times are perfectly known can be derived from the results given in~\cite{Amato:IJC11}, where also the case of state-dependent switchings has been considered.

\subsection{Time-dependent Switching Linear Systems}\label{subsection:TDSLS}
Let us consider the family of linear systems:
\begin{subequations}\label{equation:family}
\begin{align}
\dot{x}(t) & = A_p(t)x(t)+G_p(t)w(t)\,,\label{equation:familyA}\\
y(t)& =  C_p(t)x(t)\label{equation:familyB}
\end{align}
\end{subequations}
with $p\in\mathcal{P}=\bigl\{1\,,\ldots\,,l\bigr\}$, and $A_p(\cdot)~\,~:~\,~\mathbb{R}_0^+~\mapsto~\mathbb{R}^{n\times n}$,
$G_p(\cdot)\,:\,\mathbb{R}_0^+\mapsto
\mathbb{R}^{n\times r}$, and
$C_p(\cdot)\,:\,\mathbb{R}_0^+\mapsto\mathbb{R}^{m\times
n}$ are continuous matrix-valued functions.

To define a switching linear system generated by the
family~\eqref{equation:family}, the notion of \emph{switching
signal}~$\sigma(\cdot)$ is needed (see~\cite{Liberzon}). In
particular, $\sigma(\cdot):\mathbb{R}^+_0\mapsto\mathcal{P}$ is a
piecewise constant function, where the discontinuities are called
\emph{resetting times}; we denote
with~$\mathcal{T}=\bigl\{t_1\,,t_2\,,\ldots\bigr\}\subset\mathbb{R}^+_0$
the set of resetting times. Furthermore, the switching signal is
assumed to be right-continuous everywhere.

Given the family~\eqref{equation:family} and the switching signal~$\sigma(\cdot)$, the class of TD-SLS is given by
\begin{subequations}\label{equation:ImpulsiveLinearSystem}
\begin{align}
\dot{x}(t) & = A_{\sigma(t)}(t)x(t)+G_{\sigma(t)}(t)w(t)\,,\nonumber\\
\label{equation:ImpulsiveContinuous}&\hspace{2cm} x(t_0)=0\,, \quad t\not\in\mathcal{T}\\
\label{equation:ImpulsiveDiscrete}x(t^+)& = J(t)x(t)\,, \hspace{1cm} t\in\mathcal{T}\\
y(t)& =  C_{\sigma(t)}(t)x(t)
\end{align}
\end{subequations}
where $J(\cdot) \,:\, \mathbb{R}_0^+ \mapsto \mathbb{R}^{n\times n}$ is a matrix-valued function. In
particular~\eqref{equation:ImpulsiveContinuous} describes the
\emph{continuous-time} dynamics of the TD-SLS,
while~\eqref{equation:ImpulsiveDiscrete} represents the
\emph{resetting law}. The function~$\sigma(t)$ specifies, at each time instant~$t$, the linear
system currently being \emph{active}.

Note that, in the definition given above, all the linear systems in the family~\eqref{equation:family} have the same order. Although this could not be necessarily true, we make this assumption for the sake of simplicity. How to include the possibility of considering systems of different order is briefly discussed in Remark~\ref{reamrk:differentSize}.

\medskip

Without loss of generality, we assume that the first resetting time $t_1\in\mathcal{T}$ is such
that $t_1~>~t_0$. Indeed, the case $t_1 = t_0$ is equivalent to a
change of the initial state that will be in any case equal to 0, since $x(t_0)~=~0$.

\medskip

It is worth to notice that TD-SLSs include the case of time-dependent \emph{impulsive dynamical linear systems} (TD-IDLS,~\cite{Haddad:06,SSSC10,Amato:IJC11}), where a single continuous dynamic is considered. Indeed, for TD-IDLS, the switching signal $\sigma(t)$ is constant for all $t$ and can be discarded, while the resetting times set $\mathcal{T}$ corresponds to the set of times where the state jumps.

\medskip

Since we are interested in the behavior of TD-SLS in a given time interval, we assume that
\[
\Omega\cap\mathcal{T} =
\bigl\{t_1\,,t_2\,,\ldots\,,t_h\bigr\}\,,
\]
i.e., only a finite number of \emph{switches} occurs
in~\eqref{equation:ImpulsiveLinearSystem}. This also prevents the
TD-SLS~\eqref{equation:ImpulsiveLinearSystem} from exhibiting Zeno
behavior~(\cite{ZENO}).

\medskip

In the following we drop the dependency of~$\sigma$ on~$t$, in order to simplify the notation.

\subsection{IO-FTS of TD-SLS with Known Resetting Times}\label{IO-FTS FOR TD--IDLS/known_times}
We first consider the case of known resetting times
with~$\mathcal{W}_2$ input signals.

\medskip

\begin{theorem}[\cite{SSSC10} and~\cite{Carannante:MED2011}]\label{thrm:known_times_W_2}Assume that the following D/DLMI
\small
\begin{subequations}\label{eq:DLMI_2}
\begin{align}
&\begin{pmatrix} \dot{P}(t)+A^T_{\sigma}(t)P(t)+P(t)A_{\sigma}(t) & P(t)G_{\sigma}(t)\\
G^T_{\sigma}(t)P(t) & -R(t) \end{pmatrix} <0\,,\nonumber\\
 & \hspace{4cm} \forall\ t\in\Omega\,,\tau\notin\mathcal{T} \label{eq:DLMI_2_1}\\
&J^T(t_k)P(t_k^+)J(t_k)-P(t_k)\le 0\nonumber\,,\\
&\hspace{4cm}\forall\ t_k\in\Omega\cap\mathcal{T}\label{eq:DLMI_2_2}\\
\label{eq:DLMI_2_3} & P(t)>
C^T_{\sigma}(t)Q(t)C_{\sigma}(t)\,,\quad\forall\
t\in\,\Omega
\end{align}
\end{subequations}
\normalsize
admits a positive definite piecewise differentiable
matrix-valued solution~$P(\cdot)$, then
system~\eqref{equation:ImpulsiveLinearSystem} is IO-FTS with
respect to $\bigl(\mathcal{W}_2,Q(\cdot),\Omega)$.
\hfill$\blacksquare$
\end{theorem}

\medskip

\begin{remark}\label{reamrk:differentSize}
It should be noticed that when family~\eqref{equation:family} is
made of systems with different dimensions,
Theorem~\ref{thrm:known_times_W_2} still holds. Indeed, if this is
the case inequality~\eqref{eq:DLMI_2_2} can still
be defined by choosing~$P(\cdot)$ with the same dimension
of~$A_\sigma(\cdot)$ for all~$t$, and by noticing that, in
general, $J(t_k)$ is not a square matrix. \hfill$\blacktriangle$
\end{remark}

\subsection{IO-FTS of TD-SLS under Arbitrary Switching}\label{IO-FTS FOR TD--IDLS/arbitrary_switching}
The case of no knowledge of the resetting times, i.e. arbitrary
switching (AS), is tackled in this section. The main difference between the AS case and the certain case presented in the previous section is that the optimization matrix~$P(\cdot)$ cannot exhibit any jumps in~$\Omega$, since the resetting times are unknown. It turns out that in the AS case $P(\cdot)$ must be a differentiable matrix-valued function.

\medskip

\begin{theorem}[\cite{Carannante:MED2011}]\label{thrm:arbitrary switching_W_2} Assume that the following D/DLMI
\begin{subequations}\label{eq:DLMI_3}
\begin{align}
& \begin{pmatrix} \dot{P}(t)+A^T_i(t)P(t)+P(t)A_i(t) & P(t)G_i(t)\\
G^T_i(t)P(t) & -R(t) \end{pmatrix} <0 \label{eq:DLMI_3_1}\\
&J^T(t)P(t)J(t)-P(t)\le 0\label{eq:DLMI_3_2}\\
\label{eq:DLMI_3_3} & P(t)\geq C^T_i(t)Q(t)C_i(t)
\end{align}
\end{subequations}
admits a positive definite and continuous solution~$P(\cdot)$ in~$\Omega$ and for all~$i\in\mathcal{P}$, then system~\eqref{equation:ImpulsiveLinearSystem} is IO-FTS wrt~$\bigl(\mathcal{W}_2,Q(\cdot),\Omega\bigr)$ under arbitrary switching.
\hfill$\blacksquare$
\end{theorem}

\medskip

Although conditions~\eqref{eq:DLMI_3} are similar to the ones given in Theorem~\ref{thrm:known_times_W_2}, they have to be checked for each linear system in~\eqref{equation:family} in the whole time interval. This unavoidably leads to more conservatism. In particular, since $P(\cdot)$ must be continuous, inequality~\eqref{eq:DLMI_3_2} implies~$J(\cdot)$ to be Schur for all~$t$ in~$\Omega$. Hence, due to the lack of knowledge on the resetting times, it is necessary to have \emph{stable} resetting laws in order to meet conditions~\eqref{eq:DLMI_3}.

\subsection{IO-FTS of TD-SLS under Uncertain Switching}\label{IO-FTS FOR TD--IDLS/uncert_switching}
Let us now consider TD-SLS with uncertain switching (US), i.e. the case where the $k$-th resetting time is known with a given uncertainty $\pm \Delta T_k$.

Even in the US case, the sufficient condition to be checked to assess IO-FTS turns out to be more conservative with respect to the one derived in Section~\ref{IO-FTS FOR TD--IDLS/known_times}.  Furthermore a trade-off between uncertainty on the resetting times and additional constraints to be added in order to check IO-FTS clearly appears. In particular, the less is the uncertainty on the resetting times, the fewer are the additional constraints to be verified.

\medskip

Since we still consider~$\sigma(\cdot)$ piecewise constant with
discontinuities in correspondence of $t_k\in\mathcal{T}$, it is useful to introduce the following definitions to
describe the uncertainty on the resetting times
\begin{align*}
\Psi_1 & = \big]t_0\,,t_1+\Delta T_1\big[\,,\\
\Psi_j & =\big]t_{j-1}-\Delta T_{j-1}\,,t_j+\Delta T_{j}\big[\,,\, j=2\,,\ldots\,,h\\
\Psi_{h+1} & = \big]t_h-\Delta T_h\,,t_0+T\big]\\
\Gamma_j &=\big[t_{j}-\Delta T_{j},t_j+\Delta T_{j}\big]\,,
\,j=1\,,\ldots\,,h
\end{align*}
with
\begin{equation}\label{equation:condGamma}
\bigcap_{j=1}^h\Gamma_j=\emptyset\,,
\end{equation}
which implies the knowledge of the resetting times order.

\medskip

\begin{theorem}[\cite{Carannante:MED2011}]\label{thrm:uncert switching_W_2} If there exist~$h+1$ positive definite matrix-valued functions~$P_j(\cdot)$, $j~=~1\,,\ldots\,,h+1$, that satisfy the following D/DLMI
\begin{subequations}\label{eq:DLMI_4}
\begin{flalign}
& \begin{pmatrix}{\scriptstyle\dot{P_j}(t)+A^T_{\sigma(t_{j-1})}(t)P_j(t)+P_j(t)A_{\sigma(t_{j-1})}(t)} & {\scriptstyle P_j(t)G_{\sigma(t_{j-1})}(t)}\\
{\scriptstyle G^T_{\sigma(t_{j-1})}(t)P_j(t)} & {\scriptstyle -R(t)} \end{pmatrix} {\scriptstyle <0} \nonumber \\
&\hspace{3.5cm} t \in \Psi_j\,,\, j = 1\,,\ldots\,,h+1 \label{eq:DLMI_4_1}\\
&J^T(t)P_{j+1}(t)J(t)-P_{j}(t)\le 0 \,,\nonumber\\
&\hspace{3.5cm} t\in\Gamma_j\,,\, j = 1\,,\ldots\,,h\label{eq:DLMI_4_2}\\
\label{eq:DLMI_4_3} & P_j(t)\geq
C^T_{\sigma(t_j)}(t)Q(t)C_{\sigma(t_j)}(t)\,,\nonumber\\
&\hspace{3.5cm} \quad t \in \Psi_j\,,\, j = 1\,,\ldots\,,h+1
\end{flalign}
\end{subequations}
then system~\eqref{equation:ImpulsiveLinearSystem} is
IO-FTS wrt~$\bigl(\mathcal{W}_2,Q(\cdot),\Omega\bigr)$ under uncertain switching defined by~$\Delta T_k$, $k=1\,,\ldots\,,h$.
\hfill$\blacksquare$
\end{theorem}

\medskip

It should be noted that the length of $\Psi_j$ and $\Gamma_j$
decreases when the uncertainties get smaller, leading us to the
same result of Theorem~\ref{thrm:known_times_W_2} when $\Delta
T_{k}=0$ for all $k$. Eventually, note that even the partial knowledge
of the resetting times allows us to check the IO-FTS of the
TD-SLS system~\eqref{equation:ImpulsiveLinearSystem} without
requiring the different linear systems in the
family~\eqref{equation:family} to be IO-FTS.

\medskip

\begin{example}\label{example:SLS}
The following numerical example shows the effectiveness
of the proposed approach when checking IO-FTS of TD-SLS in the case of
different levels of knowledge on the resetting times. In
particular, we focus our attention on input signals belonging to the
class~$\mathcal{W}_\infty$.

Let us consider the second order TD-SLS given by the following two linear systems
\[
A_1=\left(\begin{array}{cc}0.5+\rho t & 0.1\\0.4 & -0.3+\rho
t\end{array}\right)\,,
\]
\[
G_1=\left(\begin{array}{c}1\\1\end{array}\right)\,,\,
C_1=\left(\begin{array}{cc}1& 1\end{array}\right)\,,
\]
\[
A_2=\left(\begin{array}{cc}0.15+\rho t & 0.2\\1 & -0.25-\rho
t\end{array}\right)\,,
\]
\[
G_2=\left(\begin{array}{l}1\\0\end{array}\right)\,,\,
C_2=\left(\begin{array}{ll}2& 1\end{array}\right)\,,
\]
with $\rho=0.5$.
The value of the switching signal~$\sigma(\cdot)$ in the time interval $\Omega=[0\,,1]$ is shown in Fig.~\ref{Figure:sigma}, hence
\[
[0\,,1]\cap\mathcal{T}=\bigl\{0.2\,,0.5\,,0.75\bigr\}\,.
\]
The resetting law is defined by
\[
J=\left(\begin{array}{cc}-1.1 & 0\\0 & 0.1\end{array}\right)\,.
\]
%
As for the IO-FTS we consider
\begin{equation}\label{equation:IOParams}
w(\cdot)=1\,,\, R=1\,,\, t_0=0\,,\, \mathrm{and}\ T=1\,.
\end{equation}
Note that the two considered
linear systems are both unstable. Furthermore, they are also IO finite-time unstable when considering the parameters specified in~\eqref{equation:IOParams} and when~$Q \ge 0.12$, as shown in Fig.~\ref{Figure:Diff_syst}.

\begin{figure}
\centering
\includegraphics[width=3.5in]{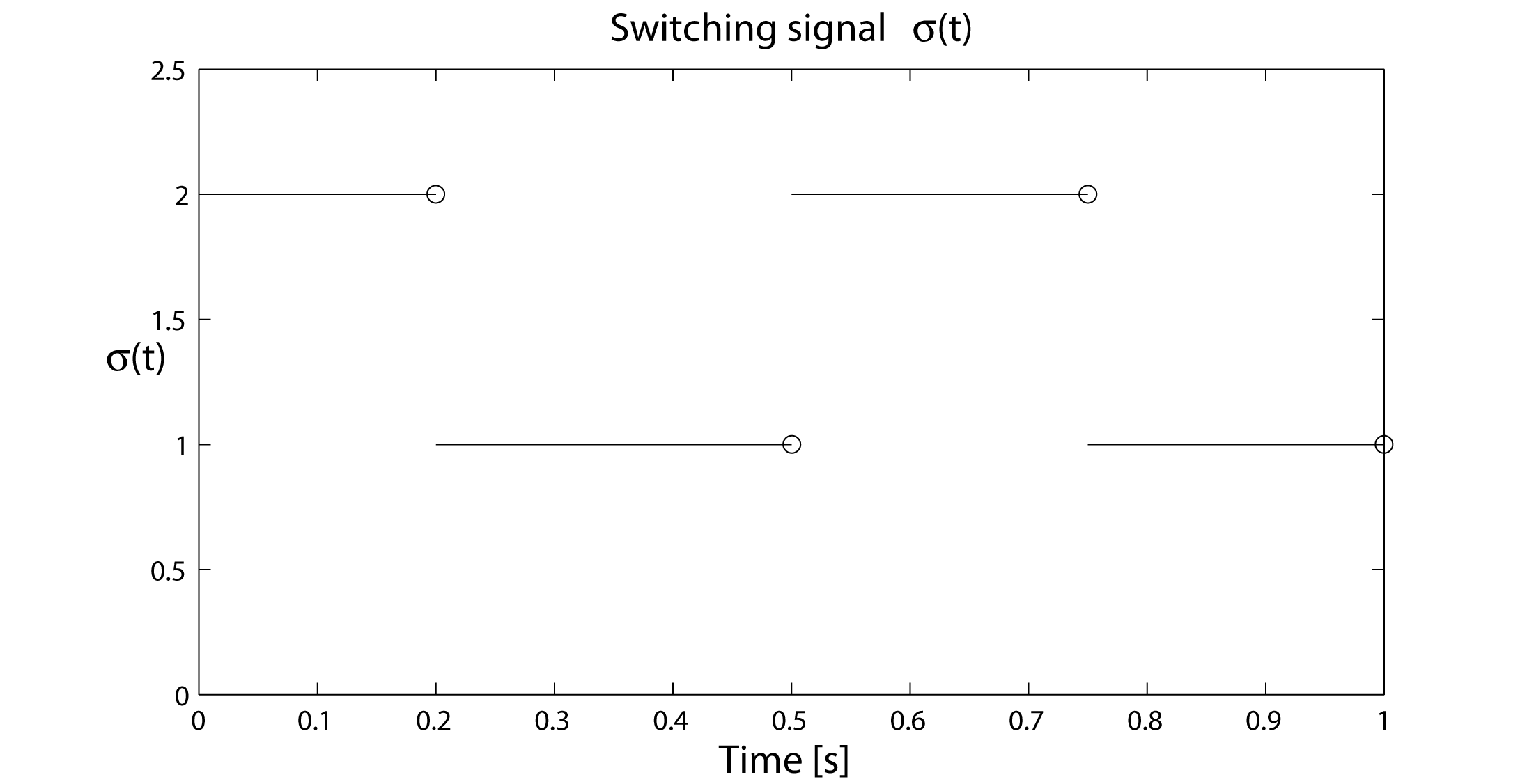}
\caption{Switching signal for the TD-SLS considered in Example~\ref{example:SLS}.} \label{Figure:sigma}
\end{figure}
\begin{figure}
\centering
\includegraphics[width=3.5in]{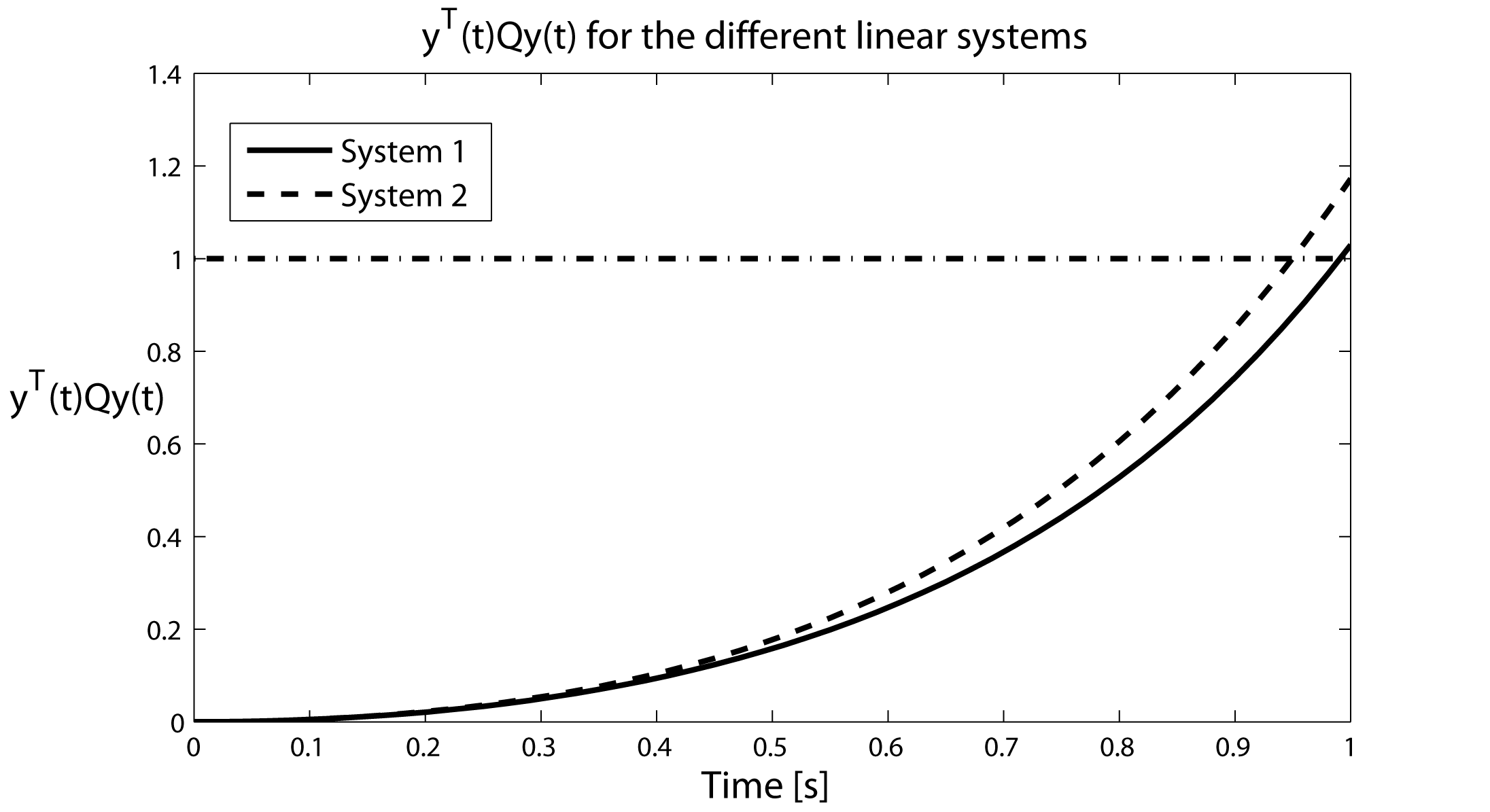}
\caption{Weighted output for the different linear systems defined in Example~\ref{example:SLS} when~$Q=0.12$.} \label{Figure:Diff_syst}
\end{figure}

\begin{figure}
\centering
\includegraphics[width=3.5in]{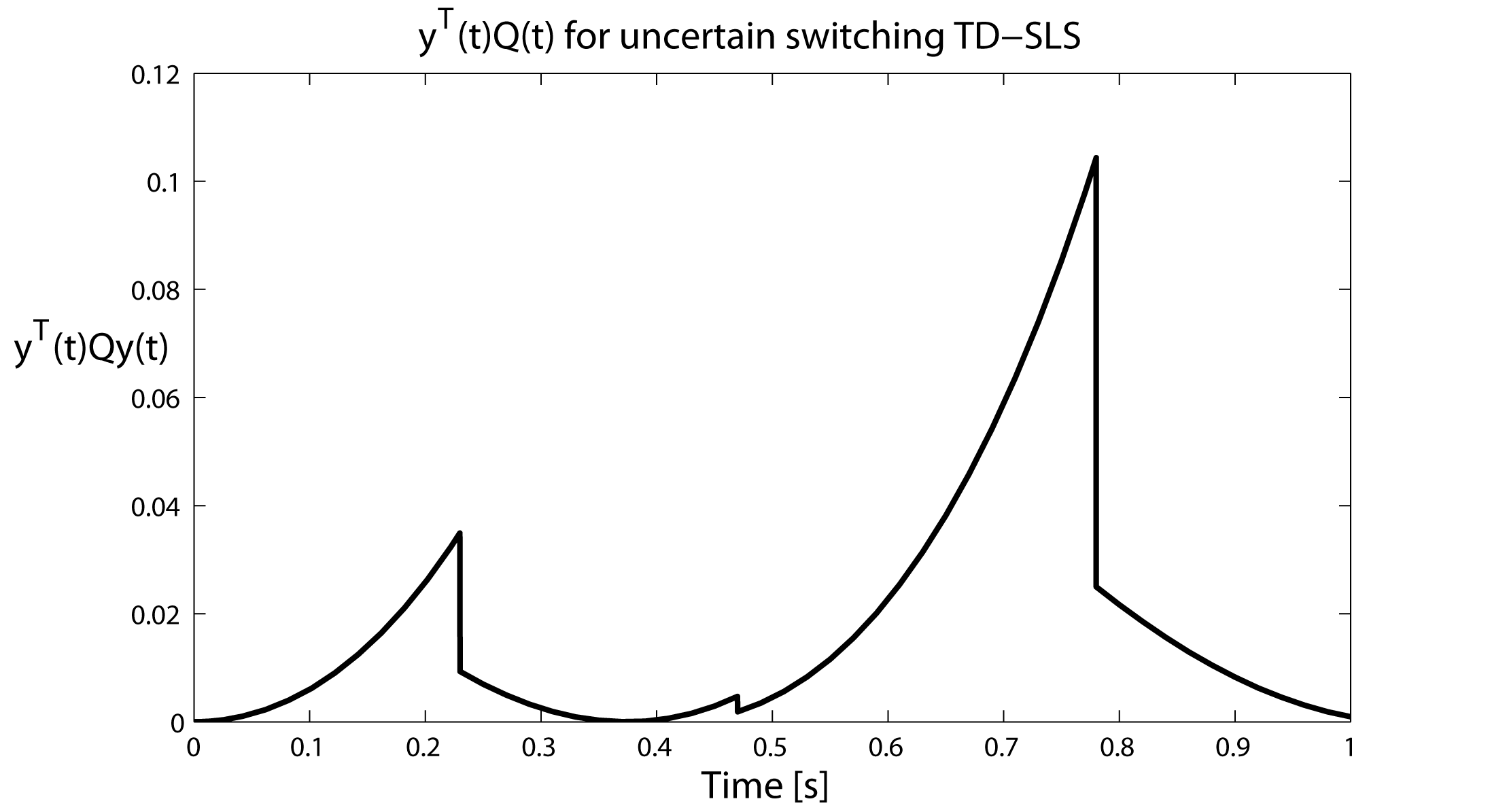}
\caption{Worst case weighted output for the uncertain switching when $Q=0.14$.} \label{Figure:US_IDLS}
\end{figure}

Let now consider the case of uncertain switching with~$\Delta T_j~=~0.03~s$  for all~$j$. The D/DLMI needed to be solved to check IO-FTS can be recast into LMIs by choosing the~$h+1$ piecewise linear matrix-valued function~$P_i(\cdot)$. By exploiting the Matlab LMI Toolbox, it turns out that the considered
system is IO-FTS wrt~$\bigl(\mathcal{W}_\infty\,,Q\,,[0\,,1]\bigr)$ for all~$Q \le 0.14$. Fig.~\ref{Figure:US_IDLS} shows
the worst case output. 
\hfill$\blacktriangle$
\end{example}

\section{EXAMPLE}\label{section:example}
\begin{figure}
      \centering
      \includegraphics[width=8cm]{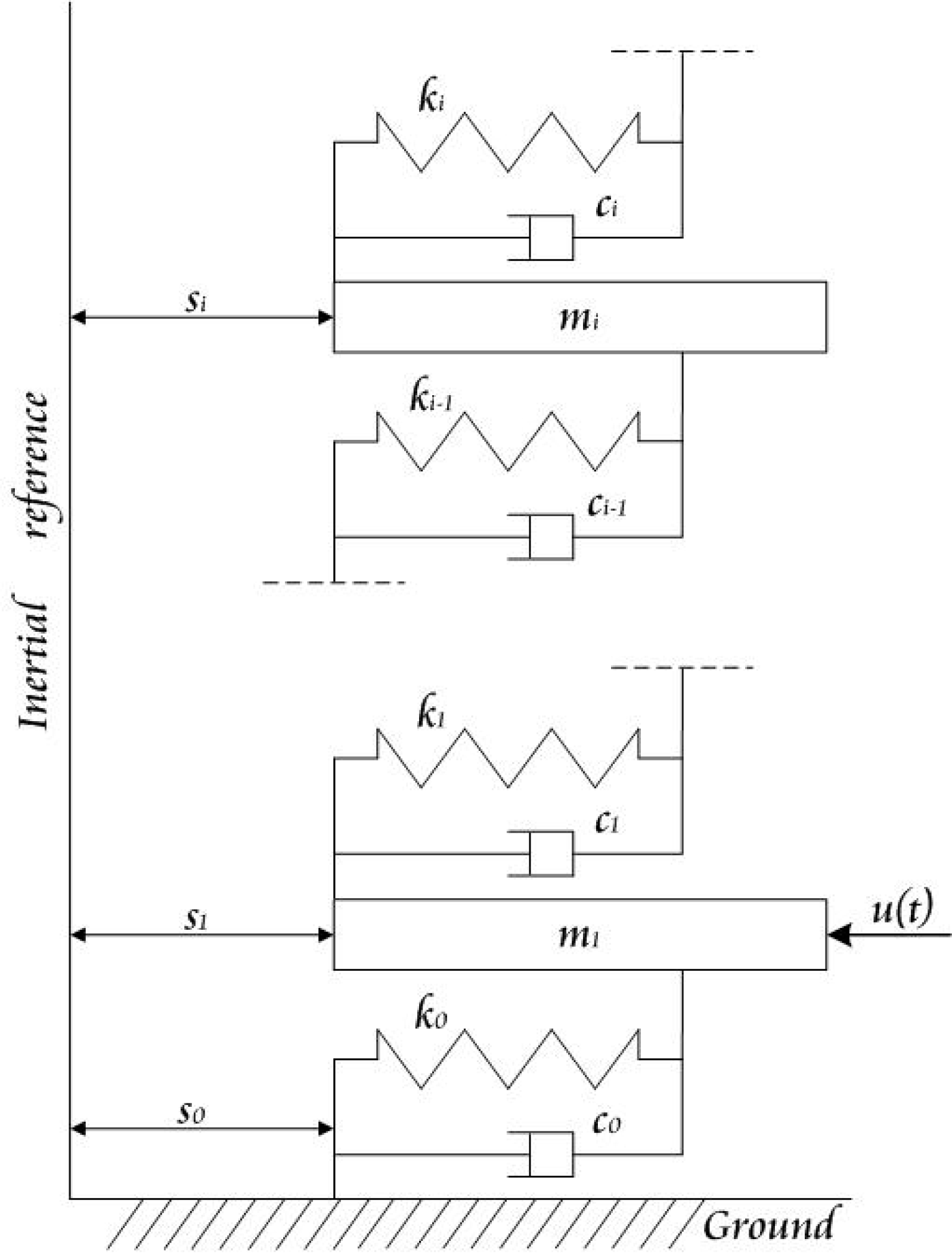}
      \caption{Lumped parameters model of a N-story building.}
      \label{figure:building}
\end{figure}

In this section we consider, as an example of application, an N-story building subject to an earthquake.
The building lumped parameters model is reported in Fig.~\ref{figure:building}.
The control system is made by a base isolator together with an actuator that generates a control force on the base floor.

The aim of the isolator is to produce a dynamic decoupling of the
structure from its foundation. If this is the case, the
inter-story drifts are reduced and the building behavior can be
approximated by the one of a rigid body~(\cite{BuildIsol}).
Furthermore, the description of the system in terms of absolute
coordinates, i.e., when the displacement is defined with respect to an
inertial reference, ensures that the disturbances act only at the
base floor~(\cite{BuildParam}).

It turns out that it is sufficient to provide an actuator only on
the base floor in order to keep the displacement and velocity of
the structure under a specified boundary. Indeed, the goal of the
control system is to overcome the forces generated by the
isolation system at the base floor, in order to minimize the
absolute displacement and velocity of the structure.

The state-space model of the considered system is

\begin{subequations}\label{equation:buildingLTI}
\begin{align}
\dot{x}(t)&=A x(t) + B u(t) + G w(t) \\
y(t)&=C x(t)
\end{align}
\end{subequations}

If we denote with~$s_0(\cdot)$ and~$\dot{s_0}(\cdot)$ the displacement and the
velocity of the ground and with~$s_i(\cdot)$ and~$\dot{s_i}(\cdot)$ the
displacement and the velocity of the i-th floor, then the state vector
can be defined as $x(\cdot) = [x_1(\cdot) \, x_2(\cdot)\, \ldots \, x_{2N}^T(\cdot)]$,
where $x_i(\cdot)=\dot{s_i}(\cdot)$ and $x_{i+N}(\cdot)=s_i(\cdot)$, $i=1\,,\ldots\,,N$. The
vector $w(\cdot)=[s_0(\cdot) \,\, \dot{s_0}^T(\cdot)]$ represents the
exogenous input and $u(t)$ is the control force applied to the
base floor. The model matrices in~\eqref{equation:buildingLTI} are equal to
\begin{equation*}\label{BuildMatrices}
\begin{array}{ll}
A=\left(\begin{array}{cc} A_1 & A_2  \\
I & 0\end{array}\right)\,, & B=\left(\begin{array}{c} 1/m_1  \\
0\end{array}\right)\,,\\ G=\left(\begin{array}{ccc} k_0/m_1  & c_0/m_1\\
0 & 0\end{array}\right)\,, &
\end{array}
\end{equation*}
\small
\begin{equation*}\label{BuildMatricesC}
C=\left(\begin{array}{cccccc} \frac{-(c_0+c_1)}{m_1} &
\frac{c_1}{m_1} & \underbrace{0\ \ldots\ 0}_{N-2} & \frac{-(k_0+k_1)}{m_1} &
\frac{k_1}{m_1} & \underbrace{0\ \ldots\ 0}_{N-2}
\end{array}\right)\,,
\end{equation*}
\normalsize
where $A_1$ and $A_2$ are $N\times N$ tridiagonal matrices defined in~\eqref{BuildMatricesA1}.
\begin{figure*}[!th]
\centering
\begin{subequations}\label{BuildMatricesA1}
\begin{align}
A_1&=\left(\begin{array}{ccccccc} -\frac{(c_0+c_1)}{m_1} & \frac{c_1}{m_1} & 0 &  & \ldots &  & 0 \\
& & & \ldots\ldots & \\
0& \ldots & \frac{c_{i-1}}{m_i} & -\frac{(c_{i-1}+c_i)}{m_i} & \frac{c_i}{m_i} & \ldots & 0\\
& & & \ldots\ldots & \\
0 & & \ldots & & 0 & \frac{c_{N-1}}{m_N} &
-\frac{c_{N-1}}{m_N}\end{array}\right)\,,\\
A_2&=\left(\begin{array}{ccccccc} -\frac{(k_0+k_1)}{m_1} & \frac{k_1}{m_1} & 0 &  & \ldots &  & 0 \\
& & & \ldots\ldots & \\
0& \ldots & \frac{k_{i-1}}{m_i} & -\frac{(k_{i-1}+k_i)}{m_i} & \frac{k_i}{m_i} & \ldots & 0\\
& & & \ldots\ldots & \\
0 & & \ldots & & 0 & \frac{k_{N-1}}{m_N} &
-\frac{k_{N-1}}{m_N}\end{array}\right)\,.
\end{align}
\end{subequations}
\line(1,0){520}
\end{figure*}

The model parameters are reported in
Table~\ref{table:Parameters} for the six story building considered in this example.

\medskip

\begin{figure*}
      \centering
      \includegraphics[width=15cm]{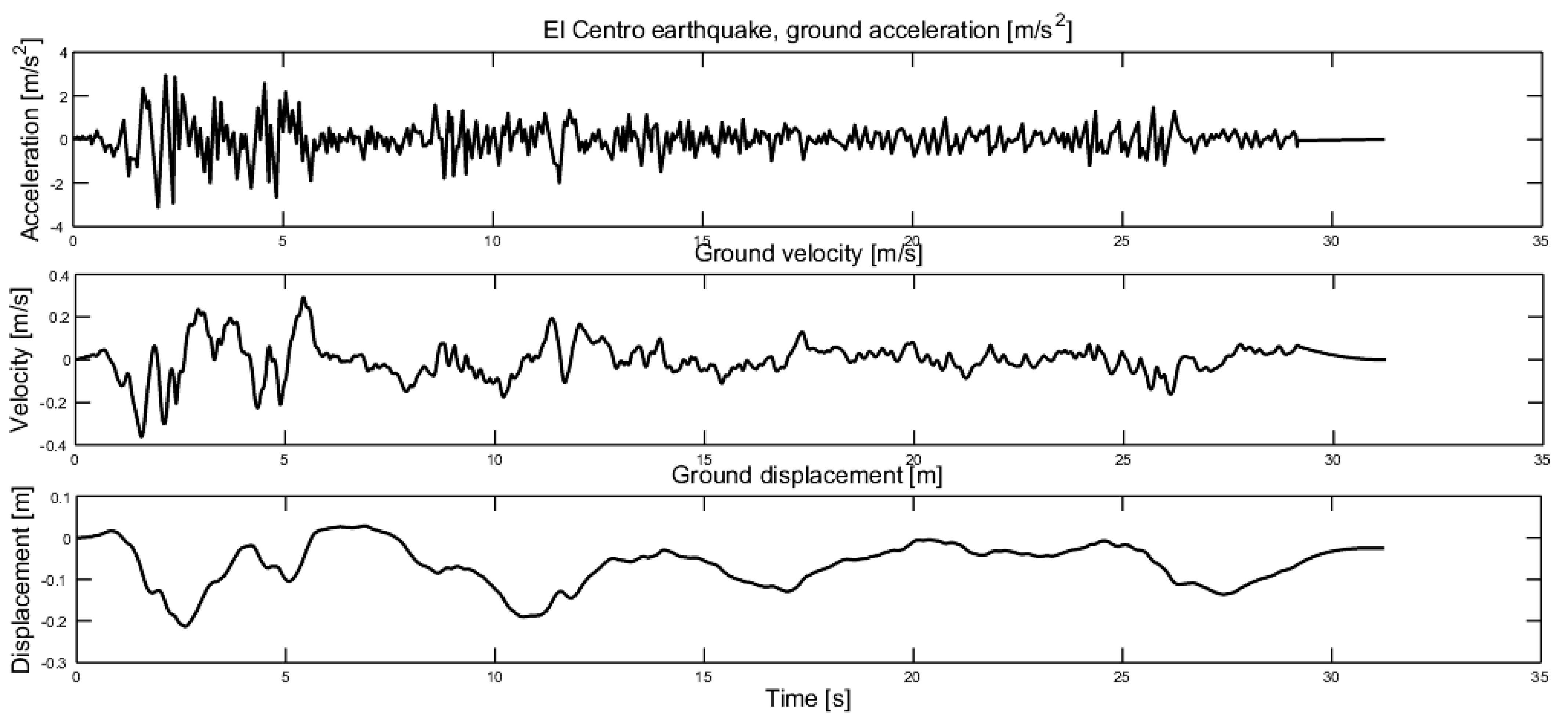}
      \caption{Ground acceleration, velocity and displacement of El Centro earthquake.}
      \label{figure:earthquake}
\end{figure*}

Taking into account the presence of the isolator and given the choice of the~$C$ matrix, the controlled output is related to the acceleration at the ground floor. Concerning the choice of the IO-FTS parameters, for a given geographic area these can be chosen starting from the worst earthquakes on record.
Indeed, from the time trace of the ground acceleration, velocity and displacement of the \emph{El~Centro}
earthquake (May 18, 1940) reported in Fig.~\ref{figure:earthquake}, the following IO-FTS parameters have been considered
\begin{equation}\label{example_parameters}
R=I\,,\,Q=0.1\,,\, \Omega=\bigl[0\,,35\bigr]\,.
\end{equation}
Exploiting Theorem~\ref{th_of} it is possible to find the controller matrix-valued functions~$A_k(\cdot)$, $B_k(\cdot)$, $C_k(\cdot)$, and~$D_k(\cdot)$ that make system~\eqref{equation:buildingLTI} IO-FTS with respect to the parameters given
in~\eqref{example_parameters}, when $\mathcal{W}_2$ disturbances are considered.

\begin{figure*}
      \centering
      \includegraphics[width=15cm]{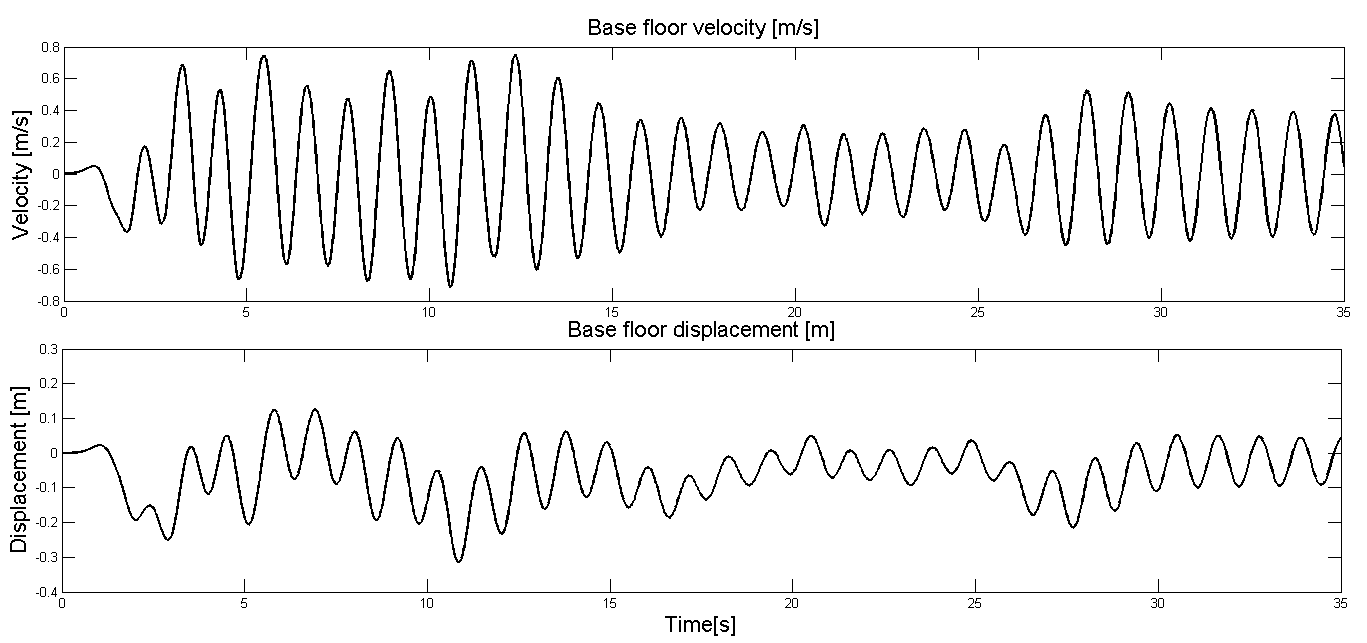}
      \caption{Uncontrolled base floor velocity and displacement.}
      \label{figure:UncontrBase}
\end{figure*}

\begin{figure*}
      \centering
      \includegraphics[width=15cm]{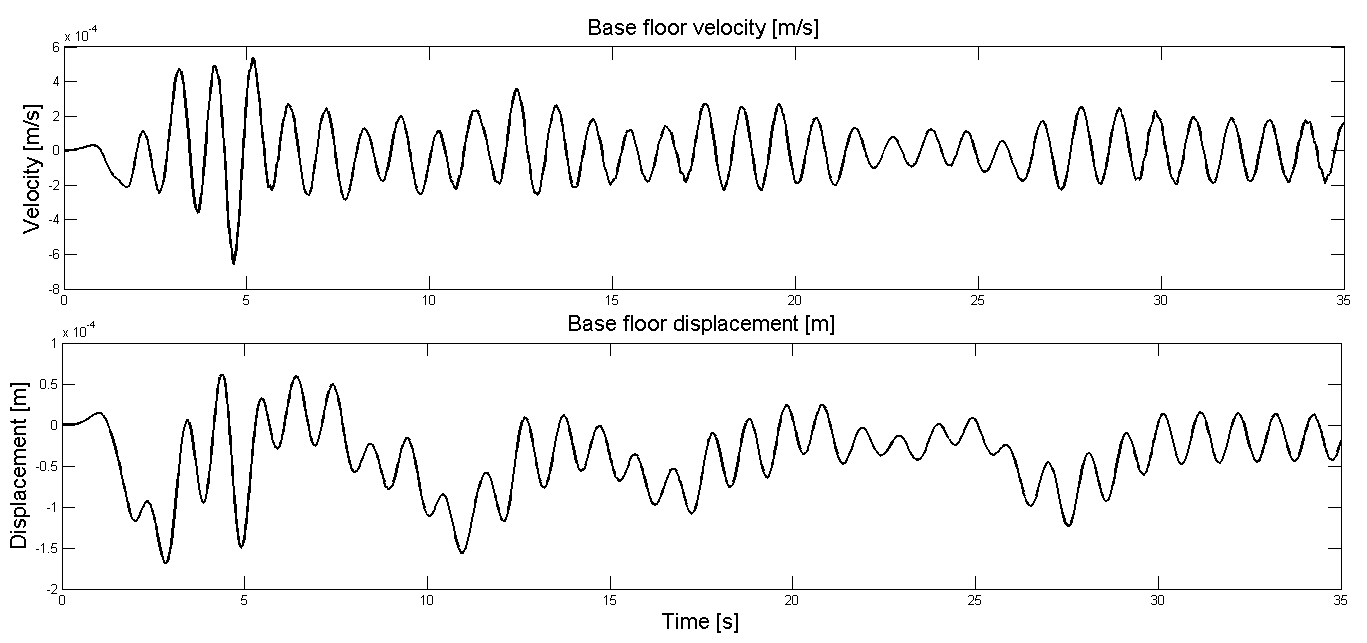}
      \caption{Controlled base floor velocity and displacement.}
      \label{figure:ContrBase}
\end{figure*}

\begin{figure*}
      \centering
      \includegraphics[width=15cm]{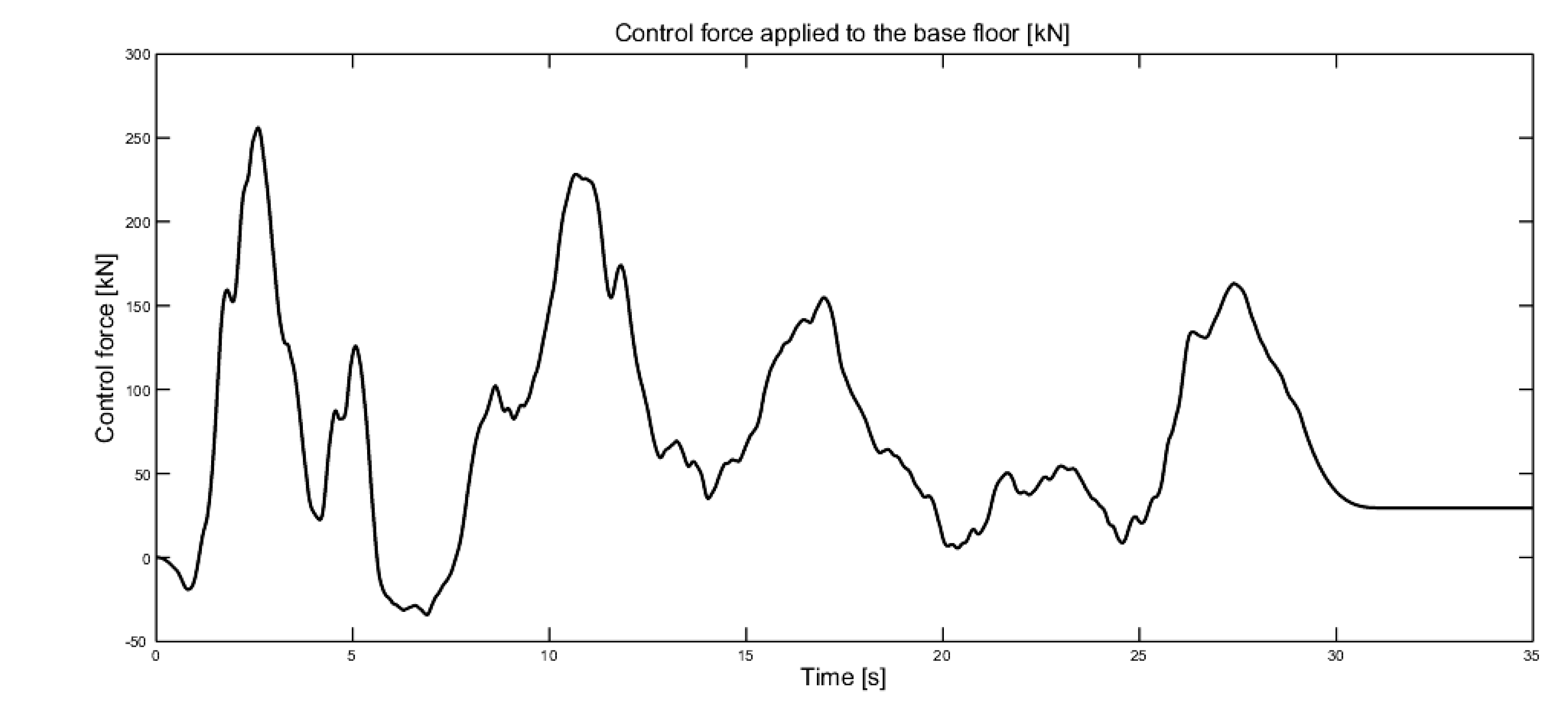}
      \caption{Control force applied to the base floor.}
      \label{figure:ForceBase}
\end{figure*}

Fig.~\ref{figure:UncontrBase} shows the base floor velocity and
displacement histories for the uncontrolled building with base
isolation system, under the assumed earthquake excitation. As it
can be seen in Fig.~\ref{figure:ContrBase}, the control system
manages to keep very small both the velocity and the displacement
of the structure. The relative control force is depicted in
Fig.~\ref{figure:ForceBase}.

\begin{table*}
\caption{Model parameters for the considered N-story building.} \label{table:Parameters}
\begin{center}
\begin{tabular}{|c||c||c|}
\hline Mass [kg] & Spring coefficient [kN/m] & Damping coefficient [kNs/m] \\ 

\hline \hline  & $k_0$=1200 & $c_0$=2.4\\ 
\hline
$m_1$=6800 & $k_1$=33732  & $c_1$=67\\
\hline\hline
$m_2$=5897 & $k_2$=29093  & $c_2$=58\\
\hline\hline
$m_3$=5897 & $k_3$=28621  & $c_3$=57\\
\hline\hline
$m_4$=5897 & $k_4$=24954  & $c_4$=50\\
\hline\hline
$m_5$=5897 & $k_5$=19059  & $c_5$=38\\
\hline\hline
$m_6$=5897 &   &  \\
\hline
\end{tabular}
\end{center}
\end{table*}
\section*{CONCLUSIONS}\label{section:Conclusions}
The concept of IO-FTS is useful to deal with the input-output behavior of dynamical linear systems,
when the focus is on the boundedness of the output signal over a finite interval of time. In this paper some recent results on IO-FTS of both linear and switching systems have been revised.

\addtolength{\textheight}{-6.4cm} 

\bibliographystyle{IEEETran}
\bibliography{IOFTS}

\end{document}